\documentclass[aps,pre,twocolumn,amssymb,nofootinbib,longbibliography,superscriptaddress]{revtex4-1}
\usepackage{multirow}
\usepackage{xcolor}
\usepackage{bm,amsmath}
\usepackage{amssymb}
\usepackage[english]{babel}
\usepackage{graphicx}
\usepackage{array}
\usepackage[normalem]{ulem}
\usepackage[dvipsnames]{xcolor}
\definecolor{darkblue}{rgb}{0,0,0.6}
\usepackage[colorlinks,linkcolor=darkblue,citecolor=darkblue,urlcolor=darkblue]{hyperref}

\begin{document}

\title{Active turbulent flows of dense persistent particles driven at constant velocity}

\author{Yonglun Jiang}

\affiliation{Department of Physics, Chengdu University of Technology, 610059 Chengdu, China}

\author{Juliane Klamser}

\affiliation{Laboratoire Charles Coulomb (L2C), Université de Montpellier \& CNRS (UMR 5221), 34095 Montpellier, France}

\author{Ludovic Berthier}

\affiliation{Gulliver, CNRS UMR 7083, ESPCI Paris, PSL Research University, 75005 Paris, France}

\date{\today}

\begin{abstract}
We study dense assemblies of self-propelled particles with infinitely persistent active forces, driven at a constant average velocity $V$ along their propulsion directions. Because the velocity $V$ introduces a finite timescale, it plays a role analogous to the shear rate in globally-deformed amorphous solids, and the system approaches a transition analogous to a yielding transition as $V \to 0$. We construct a force-velocity flow curve which is of the Herschel--Bulkley type, with a finite force threshold for flow. As $V$ decreases, the instantaneous velocity field becomes increasingly dominated by rare, localized plastic events, which produce fat tails in the velocity distribution and coexist with a slow, spatially correlated flow of streams and vortices. Many such events must accumulate throughout the system to fully relax its structure, so that the flow patterns reorganize after particles have been driven over a few particle diameters. The spatial extent of these correlated flows grows rapidly as $V$ decreases, and becomes only limited by the finite system size in our simulations. The kinetic energy spectrum displays a robust $k^{-1/2}$ regime also found in self-propelled particles at large but finite persistence times. Approaching the flow threshold thus leads to the same active turbulent state as increasing the persistence time in dense active matter, while displaying both analogies and differences with the yielding transition in sheared glassy materials. 
\end{abstract}

\maketitle

\section{Introduction}

Active matter is composed of units that consume energy to propel themselves. Examples include biological systems, such as bacterial colonies, cells and tissues~\cite{tang_cell_2024, han_local_2025}, and synthetic ones, such as self-propelled colloids, vibrated grains and robots~\cite{deseigne_vibrated_2012, lam_self-propelled_2015, briand_spontaneously_2018, zottl_emergent_2016, han_emergence_2020, scholz_rotating_2018, yang_shaping_2024, wang_hyperuniform_2025}. A hallmark of their collective behavior is active turbulence~\cite{alert_active_2022}, a chaotic flowing state with vortices and jets at scales much larger than the individual units, observed for instance in bacterial suspensions, sperm, cell monolayers and active nematics~\cite{wensink_meso-scale_2012, creppy_turbulence_2015, garcia_physics_2015, saw_topological_2017, sanchez_spontaneous_2012, guillamat_taming_2017}. Unlike inertial turbulence, it develops at vanishing Reynolds number, the flows being sustained by the energy injected at the particle scale. Active turbulence is sometimes classified as wet or dry, depending on whether momentum is conserved, and as polar or nematic, depending on the symmetry of the active units~\cite{marchetti_hydrodynamics_2013,alert_active_2022}. Its spectral properties are often non-universal, with scaling exponents that depend on the system~\cite{bratanov_new_2015,martinez-prat_scaling_2021,WeiPeng2024,alert_active_2022}. Recently, active turbulence was also found in dense assemblies of isotropic self-propelled particles, in the absence of alignment interactions or hydrodynamic coupling, the flows emerging from the competition between propulsion and steric interactions~\cite{keta_emerging_2024,KlamserBerthier2025}.

Activity profoundly changes the dynamics of dense systems, and active glass transitions have been studied extensively in models of self-propelled particles~\cite{berthier_non-equilibrium_2013, berthier2014nonequilibrium, reichhardt_active_2014, marchetti_minimal_2016, berthier_how_2017, fodor_statistical_2018, berthier_glassy_2019, mandal_random_2022, debets2022active, paul_dynamical_2023}. In such models, the active forces usually decorrelate over a finite persistence time, $\tau_p$. When $\tau_p$ is short, the dynamics resembles that of thermal glass-formers~\cite{debets2022active, KetaBerthier2022}. When $\tau_p$ becomes large, particle velocities develop spatial correlations over a length that grows with $\tau_p$~\cite{Henkes2020, CapriniMarconiPuglisi2020, SzamelFlenner2021, KetaBerthier2022, Kuroda2023, szamel2026scale}. These correlated velocities organize into mesoscale chaotic flows akin to active turbulence~\cite{keta_emerging_2024}, and for extremely persistent forces the relaxation becomes intermittent, with periods of elastic loading interrupted by bursts of plastic events~\cite{mandal_extreme_2020, keta_intermittent_2023}. Similar chaotic flows appear when dense assemblies are driven by non-reciprocal pairwise forces, suggesting that they are a generic feature of locally driven dense disordered materials~\cite{KlamserBerthier2025}.

In the limit of infinite persistence, the directions of the active forces are frozen. The system then reaches a jammed, force-balanced state when their amplitude is below a threshold, while it flows above~\cite{LiaoXu2018, mo_rheological_2020, mandal_extreme_2020, villarroel_critical_2021, villarroel_avalanche_2024}. This behavior is reminiscent of the yielding of amorphous solids under shear, a correspondence that has been explored at several levels. Numerical studies showed that dense self-propelled particles display yielding and shear-thinning flow curves analogous to those of sheared materials~\cite{mo_rheological_2020, mo_thinning_2024}. Locally and globally driven systems can be connected continuously by tuning the spatial correlation of the driving forces, with equivalent avalanche statistics in the quasistatic limit~\cite{morse_manning_2021, Agoritsas_2021,agoritsas_morse_2024, villarroel_critical_2021, villarroel_avalanche_2024}. Whether the critical properties at the threshold are the same for local and global driving remains debated, as different flow exponents have been reported in some regimes~\cite{villarroel_critical_2021, villarroel_avalanche_2024, relmucao-leiva_bridging_2025}. Above the force threshold, recent simulations at fixed force amplitude found velocity correlation lengths that grow with system size without saturating, together with a segregation of particles with similar propulsion directions and the formation of large density voids~\cite{Dutta2025, Shukla2025}.

These results raise the question of how the flowing state of persistent active matter approaches the flow threshold. At finite $\tau_p$, the persistence time provides the timescale that controls the size of the correlated flows, and the turbulent state is reached by increasing $\tau_p$. In sheared amorphous solids, the approach to yielding is instead controlled by the strain rate $\dot\gamma$, which sets the timescale $1/\dot\gamma$ of the deformation, so that the quasistatic limit can be approached in a controlled way. Here we follow the same strategy for persistent active matter. Following Ref.~\cite{mo_rheological_2020}, we impose the average velocity $V$ of the particles along their propulsion directions, rather than the amplitude of the active forces, a protocol analogous to shearing at constant shear rate. The forces can then adjust their average amplitude, $F$, but remain infinitely persistent in direction. Both choices are equivalent in the thermodynamic limit~\cite{mo_rheological_2020}, and imposing $V$ simply gives direct control over the approach to the flow threshold. The limit $V \to 0$ then continuously brings the system closer to the flow threshold.

We find that the flow curve $F=F(V)$ is of Herschel--Bulkley type, with a finite yield force, $F_{\mathrm{c}}$, and an exponent $n \simeq 1/2$ reminiscent of sheared amorphous solids. At any given time, only a small fraction of the particles undergo localized plastic events. At small $V$, these rare and fast events nevertheless dominate the velocity fluctuations, giving fat tails to the velocity distribution and a rapid decay of velocity correlations on a microscopic timescale that depends only weakly on $V$. They take place on top of a slow, spatially correlated flow made of streams and vortices. Relaxing the structure requires many events to accumulate throughout the system, which happens once particles have been driven over a few particle diameters. The flow patterns reorganize on this scale, so that their lifetime scales as $1/V$. The size of these flows grows as $V$ decreases, but remains limited by the system size in our simulations. Their energy spectrum displays a $k^{-1/2}$ regime also found at finite persistence~\cite{keta_emerging_2024} and with non-reciprocal driving~\cite{KlamserBerthier2025}. The driving rate thus sets the timescale of the flow, as the shear rate does in sheared amorphous solids, and the system reaches a turbulent state similar to that of dense active matter with large but finite persistence.

The paper is organized as follows. Section~\ref{sec:model} defines the model and the numerical methods. Section~\ref{sec:macroscopic} presents the flow curve. Section~\ref{sec:single} analyzes single-particle velocity statistics and displacements. Section~\ref{sec:collective} characterizes the collective flows. Section~\ref{sec:conclusion} summarizes our results and discusses some perspectives.

\section{Model and numerical details}

\label{sec:model}

We study $N$ athermal self-propelled particles in a square box of linear size $L$, with periodic boundary conditions. The particle positions $\mathbf{r}_i$ follow an overdamped dynamics
\begin{equation}
    \chi\, \dot{\mathbf{r}}_i = -\sum_{j\ne i}\nabla_iU(r_{ij}) + F(t)\, \hat{\mathbf{n}}_i\,,
    \label{eq:overdamped}
\end{equation}
where $\chi$ is the damping coefficient, $U(r_{ij})$ the pairwise interaction potential, with $r_{ij} = |\mathbf{r}_i-\mathbf{r}_j|$, and $F(t)\hat{\mathbf{n}}_i$ is the self-propulsion force of magnitude $F(t)$ acting along a unit vector $\hat{\mathbf{n}}_i$ attached to particle $i$. Particles interact via a soft repulsive potential~\cite{ninarello_models_2017},
\begin{equation}
    U(r_{ij})=\epsilon\left(\frac{\sigma_{ij}}{r_{ij}}\right)^{12}+c_0+c_2\left(\frac{r_{ij}}{\sigma_{ij}}\right)^2+c_4\left(\frac{r_{ij}}{\sigma_{ij}}\right)^4
    \label{eq:potential}
\end{equation}
for $r_{ij}/\sigma_{ij}\leq x_c = 1.25$ and $U = 0$ otherwise, where $\epsilon$ sets the energy scale, $\sigma_{ij}$ sets the interaction range, and the constants $c_0=-28\epsilon/x^{12}_{c}$, $c_2=48\epsilon/x^{14}_{c}$, and $c_4=-21\epsilon/x^{16}_{c}$ ensure that the potential and its first two derivatives vanish continuously at the cut-off~\cite{berthier_efficient_2019}. To frustrate positional order, particle diameters $\sigma_i$ are drawn from a polydisperse distribution $P(\sigma_i)=A/\sigma_i^{3}$ with $0.73<\sigma_i<1.62$ and mean diameter $\sigma = \overline{\sigma_i}=1$, and the interaction range $\sigma_{ij}$ is taken non-additive, $\sigma_{ij} =0.5(\sigma_i+\sigma_j)(1-0.2|\sigma_i-\sigma_j|)$. The unit length is $\sigma$, the unit energy is $\epsilon$, and the unit time is $\chi\sigma^2/\epsilon$. The self-propulsion directions $\hat{\mathbf{n}}_i$ are randomly oriented with a global constraint $\sum_i \hat{\mathbf{n}}_i = 0$ to eliminate net force effects. These orientations are permanently frozen.

Following Ref.~\cite{mo_rheological_2020}, the average velocity along the self-propulsion orientation,
\begin{equation}
    V=\frac{1}{N}\sum_{i = 1}^N \dot{\mathbf{r}}_i \cdot \hat{\mathbf{n}}_i\,,
\end{equation}
is chosen as the control parameter and kept constant. This constraint makes the self-propulsion amplitude $F(t)$ dynamical, leading to fluctuations in time and across configurations. We denote by $F$ its time average in the steady state.
Projecting Eq.~(\ref{eq:overdamped}) onto $\hat{\mathbf{n}}_i$ and averaging over particles yields
\begin{equation}
\label{eq:f}
    F(t) = \chi V - \frac{1}{N}\sum_{i=1}^{N} \mathbf{f}_i \cdot \hat{\mathbf{n}}_i\,,
\end{equation}
where $\mathbf{f}_i = -\sum_{j\ne i}\nabla_iU(r_{ij})$ is the total steric force acting on particle $i$. In practice, $F(t)$ is adjusted at each integration step according to Eq.~(\ref{eq:f}) to hold $V$ to a desired constant value. We define $V$ as our driving rate, similarly to the shear rate in slowly sheared glasses.

We simulate systems with $N = 2 \times 10^3$, $10^4$, and $64 \times 10^3$ particles at fixed number density $\rho = 1$, corresponding to linear sizes $L = 45$, $100$, and $253$, respectively. We study driving rates $V = 10^{-4}$, $10^{-3}$, $10^{-2}$ and $10^{-1}$, and the flow curve in Sec.~\ref{sec:macroscopic} also includes $V = 10^{-5}$ for $N = 2 \times 10^3$. The equations of motion are integrated with the Euler method, using time steps $dt = 4 \times 10^{-3}$, $4 \times 10^{-3}$, $10^{-3}$, and $5 \times 10^{-4}$ for $V = 10^{-4}$, $10^{-3}$, $10^{-2}$, and $10^{-1}$, respectively. Initial configurations are obtained by instantaneously quenching to zero temperature liquid configurations equilibrated at a finite temperature. For each of ten independent positional configurations, we assign ten independent sets of orientations $\{\hat{\mathbf{n}}_i\}$. Since $\sum_i \hat{\mathbf{n}}_i = 0$ and interaction forces sum to zero, the total velocity vanishes at all times and the system has no net flow. We checked that results do not depend on the specific choice of the $\hat{\mathbf{n}}_i$. For $N = 64 \times 10^{3}$, we use five positional configurations, each combined with five orientation sets. We do not include results of our study for collective observables for the highest velocity $V = 10^{-1}$, because we observe the formation of empty voids, which are more prominent in larger systems, as also observed at fixed propulsion force~\cite{Dutta2025}. We checked that voids do not affect the statistics of single-particle quantities, and we include some results for $V=10^{-1}$ for those observables. 

Each run is driven over a distance $d = Vt$. We consider that the system has reached a steady state once $F(t)$ fluctuates around a constant mean value (not shown), a criterion similar to the one used for sheared systems. Because of computational cost, the steady-state window, measured in driving distance, shrinks with increasing system size and decreasing $V$, for instance $d \in [1010, 2010]$ for $N = 10^4$ at $V = 10^{-2}$, and $d \in [13, 17]$ for $N = 64 \times 10^3$ at $V = 10^{-4}$. We checked that results do not depend on the chosen window by comparing successive windows, for instance $d \in [1010, 2010]$ and $[2010, 3010]$ for $N = 10^4$ at $V = 10^{-2}$, and $d \in [13, 15]$ and $[15, 17]$ for $N = 64 \times 10^3$ at $V = 10^{-4}$. Throughout the paper, $\langle \cdots \rangle$ denotes an average over particles, steady-state times, and independent configurations.

\section{Macroscopic force-velocity response}

\label{sec:macroscopic}

\begin{figure}
    \centering
    \includegraphics{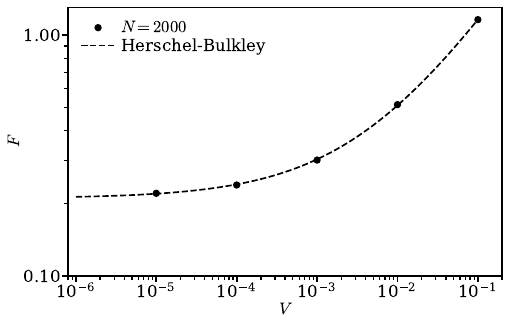}
    \caption{Macroscopic active flow curve, showing the average propulsion force $F$ as a function of the imposed driving rate $V$ for $N = 2 \times 10^{3}$ particles (symbols). The dashed line represents the Herschel--Bulkley fit [Eq.~\eqref{eq:herschel_bulkley}] with $F_{\mathrm{c}} = 0.21$, $K = 3.0$, $n = 0.5$.}
    \label{fig:HB}
\end{figure}

To characterize the collective response of our system, we first examine the macroscopic relation between the imposed driving rate $V$ and the resulting average propulsion force $F$. This ``flow curve'' $F=F(V)$ quantifies the rheological response of the system and has been shown to coincide, in the thermodynamic limit, with the one obtained when the propulsion force is fixed instead of $V$~\cite{mo_rheological_2020}.

When $V$ is imposed, the flow curve reveals a finite force threshold as $V \to 0$ (Fig.~\ref{fig:HB}). A finite propulsion force $F_{\mathrm{c}}>0$ is thus required to sustain steady motion, as for the yield stress of sheared amorphous solids. In this analogy, $V$ plays the role of the strain rate $\dot\gamma$ and $F$ that of the shear stress, and the two forms of driving can be continuously connected by tuning the spatial correlation of the driving forces~\cite{morse_manning_2021, Agoritsas_2021}. The flow curve is well described over four decades of $V$ by a Herschel--Bulkley form~\cite{Bonn2017},
\begin{equation}
F = F_{\mathrm{c}} + K V^n ,
\label{eq:herschel_bulkley}
\end{equation}
where $F_{\mathrm{c}}$ is the force threshold, $K$ a prefactor and $n$ the flow exponent. We find $F_{\mathrm{c}} = 0.21$, $K = 3.0$ and $n = 0.5$. This exponent is typical of flow curves measured in a broad range of sheared amorphous materials~\cite{Bonn2017}, and was also reported for the same model~\cite{mo_rheological_2020}.

Reducing $V$ thus brings the system toward the flow threshold $F_{\mathrm{c}}$. In sheared amorphous solids, the approach to the yield stress is controlled by the timescale $1/\dot\gamma$ over which the material is deformed. Here the driving rate introduces the analogous timescale $\sigma/V$. In the following, we show that this timescale controls the velocity statistics, and that the correlated flows grow as $V \to 0$, reminiscent of the effect of increasing the persistence time in self-propelled particles. 

\section{Single particle observables}

\label{sec:single}

We now turn to the single-particle dynamics underlying the flow curve, and characterize the velocity statistics, the temporal correlations of velocities, and particle mean-squared displacements.

\subsection{Velocity distributions}

\begin{figure}
    \centering
    \includegraphics{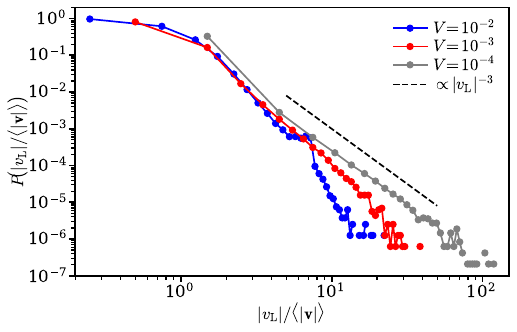}
    \caption{Distribution of the longitudinal velocity $|v_{\mathrm{L}}|/\langle |\mathbf{v}|\rangle$ for $N = 64 \times 10^3$ at different driving rates $V$. The dashed line indicates $P \propto v_{\mathrm L}^{-3}$.}
    \label{fig:pdfv}
\end{figure}

To characterize the fluctuations of the steady-state flow, we examine the stationary distribution of single-particle velocities. For each particle, the instantaneous velocity $\mathbf{v}_i = \dot{\mathbf{r}}_i$ is decomposed into components longitudinal ($v_{\mathrm{L}}$) and transverse ($v_{\mathrm{T}}$) to its propulsion direction $\hat{\mathbf{n}}_i$.

Although the driving acts only along $\hat{\mathbf{n}}_i$, the longitudinal and transverse components have indistinguishable, symmetric distributions (not shown). Multiplying Eq.~(\ref{eq:overdamped}) by $\mathbf{v}_i$ and summing over particles gives $\chi\sum_i|\mathbf{v}_i|^2 = -\mathrm{d}U_{\rm tot}/\mathrm{d}t + N F(t)\,V$, where $U_{\rm tot}=\sum_{i<j}U(r_{ij})$ is the total potential energy and we used $\sum_i \mathbf{v}_i\cdot\hat{\mathbf{n}}_i = NV$. In the steady state, $U_{\rm tot}$ fluctuates around a constant value, so that its time derivative averages to zero, and
\begin{equation}
\chi\,\langle|\mathbf{v}|^2\rangle = F\,V\,.
\label{eq:energy}
\end{equation}
This relates the velocity fluctuations directly to the flow curve. In particular, $\langle|\mathbf{v}|^2\rangle/V^2 = F/(\chi V)$ diverges as $V \to 0$, since $F \to F_{\mathrm{c}} > 0$. The drift $V$ is thus negligible compared to typical velocities, and collisions make the fluctuations isotropic.

Since the fluctuations are symmetric and isotropic, we show only the distribution $P(|v_{\mathrm{L}}|/\langle |\mathbf{v}| \rangle)$ in Fig.~\ref{fig:pdfv}. It is strongly non-Gaussian at all driving rates, with broad tails that decay more slowly than exponentially and become increasingly prominent as $V$ decreases. Similar fat tails are observed in athermal disordered systems under shear, where they originate from plastic events~\cite{LemaitreCaroli2007,Heussinger2010}. In two dimensions, the $1/r$ elastic far field of such events~\cite{PicardBocquet2004,Tanguy2006} predicts $P(v) \sim v^{-3}$ at large $v$, which is compatible with our data at the smallest driving rate. 

Together with the growing weight of fast relaxations in the velocity autocorrelation (Sec.~\ref{vv}), this indicates that as the flow threshold is approached, an increasing fraction of particle motion occurs through localized plastic events superimposed on the slow collective flow, and at any moment in time, a very small fraction of particles is actually moving much faster than the average. This physics is obviously reminiscent of slowly deformed amorphous solids~\cite{berthier2025yielding}.

\subsection{Velocity autocorrelation function}

\label{vv}

\begin{figure}
    \centering
    \includegraphics{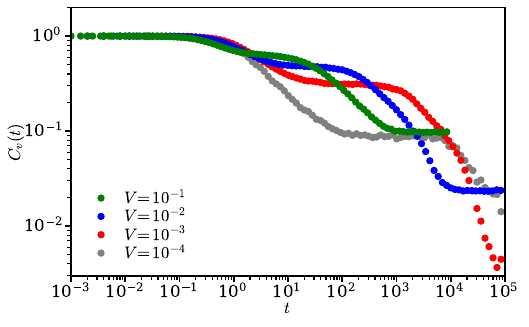}
    \caption{Velocity autocorrelation function $C_v(t)$ for $N = 10^4$ at different driving rates $V$.}
    \label{fig:vv}
\end{figure}

We further characterize particle dynamics via the velocity autocorrelation function
\begin{equation}
C_v(t) = \frac{\langle \mathbf{v}_i(0) \cdot \mathbf{v}_i(t) \rangle}{\langle |\mathbf{v}|^2 \rangle}.
\end{equation} 

Figure~\ref{fig:vv} shows that $C_v(t)$ decays in two steps governed by distinct clocks. The first decay occurs at $t \approx 1$--$10$ for all driving rates. This corresponds roughly to the time $t \approx \chi \sigma/ F$ needed for a particle to move over a particle diameter under a force of the order of the average propulsion force. Near the flow threshold, $F$ remains close to $F_{\mathrm{c}}$ (Fig.~\ref{fig:HB}), giving $\chi\sigma/F \approx 4$ at $V = 10^{-4}$, whereas the larger force at $V = 10^{-1}$ gives $\chi\sigma/F \approx 1$, consistent with the slightly earlier decay observed there. This first decay thus reflects the intrinsic duration of local rearrangements set by the overdamped dynamics, rather than the driving. As in slowly sheared amorphous solids, the driving controls how often such rearrangements occur, but only weakly affects how long each one lasts.

Since $C_v(0)=1$ represents the total velocity variance, the level reached after this first decay measures the fraction of $\langle |\mathbf{v}|^2 \rangle$ carried by motion that remains correlated over longer times. This fraction drops from about $0.6$ at $V = 10^{-1}$ to about $0.1$ at $V = 10^{-4}$. At small $V$, about $90\%$ of the velocity variance therefore decorrelates on microscopic timescales. Because $C_v(t)$ weighs each particle by its squared velocity, this does not imply that most particles move incoherently. Instead, together with the fat tails in Fig.~\ref{fig:pdfv}, it shows that the velocity fluctuations are dominated by a small fraction of fast particles undergoing short-lived plastic events, while most particles follow the slow collective flow.

The second decay marks the loss of correlation of the collective flow patterns. At all $V$, it occurs once particles have been driven over a distance $Vt$ of a few particle diameters, so the corresponding time $\tau_v$ scales approximately as $1/V$. Indeed, when $C_v$ is plotted as a function of the driving distance $Vt$ (not shown), the second decays of all driving rates fall in the same range of $Vt$. Since only a few particles undergo plastic events at any given time, reorganizing the flow requires many events to accumulate throughout the system, and their number is controlled by the driving distance $Vt$ rather than by time. This is the finite timescale introduced by the driving at constant velocity, analogous to the inverse strain rate in sheared amorphous solids. 

Finally, $C_v(t)$ reaches a long-time plateau because each particle drifts on average with velocity $V$ along its own propulsion direction, so that $C_v(t \to \infty) = V^2/\langle |\mathbf{v}|^2 \rangle = \chi V/F$, using Eq.~(\ref{eq:energy}). This plateau therefore vanishes linearly as $V \to 0$, since $F \to F_{\mathrm{c}}$. For the driving rates where the plateau is reached within our time window, its value agrees with $\chi V/F$ computed from the flow curve.

\subsection{Mean-squared displacements}

\label{msd}

\begin{figure}
    \centering
    \includegraphics{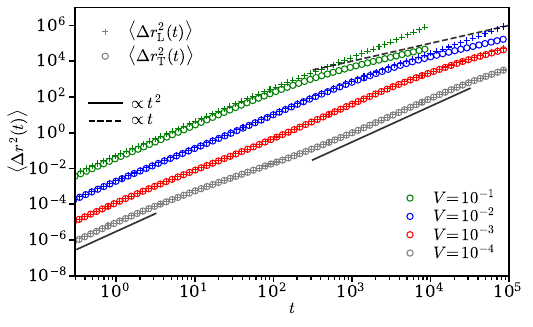}
    \caption{Longitudinal (plus) and transverse (circles) mean-squared displacements for $N = 10^4$ at different driving rates $V$. Solid and dashed lines indicate ballistic and diffusive behavior, respectively.}
    \label{fig:msd}
\end{figure}

Despite its formal relation to the velocity autocorrelation function, the mean-squared displacement (MSD) provides a complementary probe of particle transport. We decompose it into components longitudinal and transverse to the propulsion direction $\hat{\mathbf{n}}_i$:
\begin{equation}
\begin{aligned}
    \langle \Delta r^2_{L}(t) \rangle &= \langle (\Delta \mathbf{r}_i \cdot \hat{\mathbf{n}}_i)^2 \rangle\,,\\
    \langle \Delta r^2_{T}(t) \rangle &= \langle |\Delta \mathbf{r}_i - (\Delta \mathbf{r}_i \cdot \hat{\mathbf{n}}_i)\hat{\mathbf{n}}_i|^2 \rangle\, ,
\end{aligned}
\end{equation}
with $\Delta \mathbf{r}_i(t)$ the displacement of particle $i$ over a time interval $t$. Figure~\ref{fig:msd} shows that at short times both components grow as $t^2$. Since Eq.~(\ref{eq:overdamped}) has no inertia, this ballistic regime reflects the persistence of particle velocities over the microscopic relaxation time identified in Fig.~\ref{fig:vv}. 

Up to $\tau_v$ the two components are nearly identical, since velocity fluctuations are much larger than $V$ [Eq.~(\ref{eq:energy})]. At long times they become different. The longitudinal MSD becomes ballistic, $\langle \Delta r^{\,2}_{\mathrm{L}} \rangle \approx (Vt)^2$, because each particle drifts with average velocity $V$ along $\hat{\mathbf{n}}_i$, while the transverse MSD instead becomes diffusive. This separation occurs once the drift overtakes the fluctuating contribution, which happens within our time window only for $V \geq 10^{-2}$. In sheared materials, one usually separates the affine advection along the flow from the non-affine motion. Here the drift $Vt$ along $\hat{\mathbf{n}}_i$ plays the role of the affine displacement, and the transverse motion probes the non-affine dynamics.

\begin{figure*}
    \centering
    \includegraphics{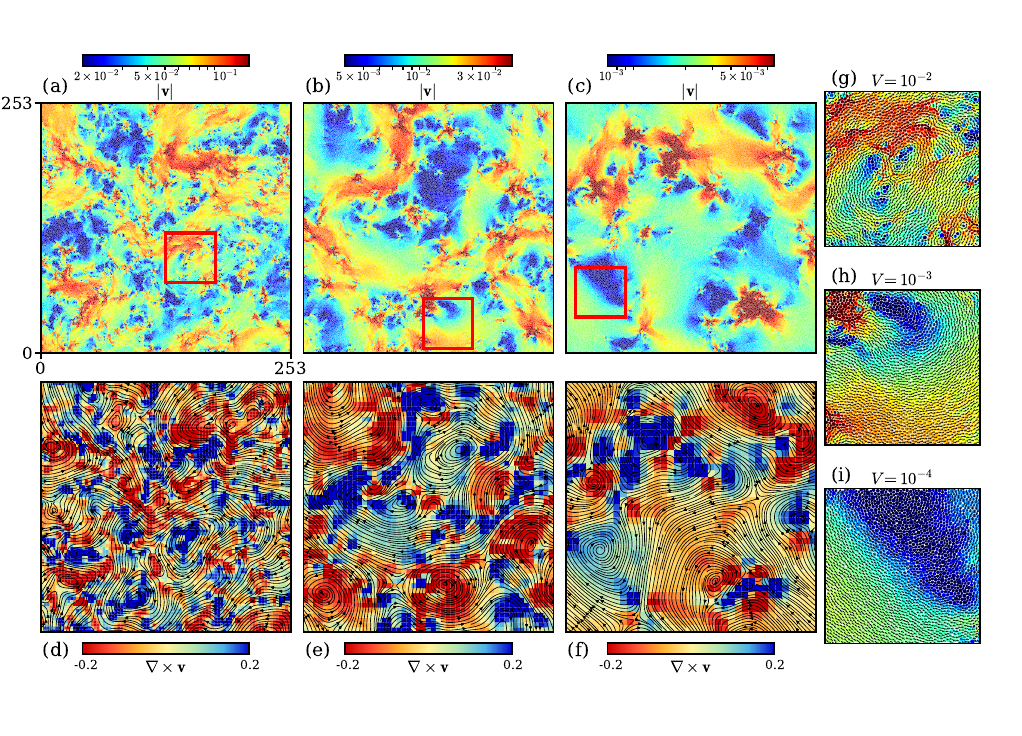}
    \caption{Snapshots of the velocity field for $N = 64000$ at $V = 10^{-2}$, $10^{-3}$ and $10^{-4}$ (left to right). (a-c) Particle configurations, with particles colored by their speed $|\mathbf{v}|$. (d-f) Vorticity of the coarse-grained velocity field, with streamlines. (g-i) Zoom on the regions marked in (a-c), with the same coloring, and arrows indicating the direction of particle velocities.}
    \label{fig:snapshot}
\end{figure*}

\section{Collective dynamics}

\label{sec:collective}

We now turn to the collective dynamics underlying the single-particle behavior discussed above, and characterize the spatial organization and correlations of the velocity field. The self-propulsion orientations are initially uncorrelated in space. In the steady state reached at long enough times, we detect only a weak growth of their spatial correlations, much weaker than for the velocities, and no clear coarsening of domains of aligned forces over the duration of our simulations. A segregation of particles with similar orientations has recently been reported in dense active fluids at fixed propulsion force, in a regime where the density field also develops large voids~\cite{Dutta2025, Shukla2025}. Such sorting requires particles to travel distances comparable to the domain size, so its development may depend on the order in which the limits of long times, $t \to \infty$, and large system sizes, $L\to\infty$ are taken. The velocity correlations discussed below are therefore not simply inherited from correlated driving forces, but emerge from the interplay of self-propulsion and crowding at high density.

\subsection{Direct visualization of the turbulent flow}

\label{snapshot}

Before quantifying spatial correlations, we begin with direct visualizations of the velocity field, which provide an intuitive illustration of the emergence and growth of correlated flow structures upon approaching the flow threshold. In the steady state, the instantaneous velocity field is highly heterogeneous (Fig.~\ref{fig:snapshot}a--c). At large $V$, fast regions of various sizes are spread throughout the system, whereas at small $V$ they become increasingly sparse and localized, consistent with the fat tails of the velocity distribution and with the growing weight of short-lived plastic events in $C_v(t)$. These localized events take place on top of a slow collective flow, revealed by the vorticity field and the streamlines of the coarse-grained velocity, which form interconnected streams and vortices whose size grows as $V$ decreases, the largest ones becoming comparable to the system size at the smallest $V$ (Fig.~\ref{fig:snapshot}d--f). The coarse-grained velocity field is obtained from  Gaussian-weighted sums of the particle velocities, with kernel standard deviations $\ell \simeq 2.98\sigma$, $4.96\sigma$, and $6.84\sigma$ for $V=10^{-2}$, $10^{-3}$, and $10^{-4}$, respectively. The vorticity is computed from the curl of this field and normalized by its maximum absolute value in each snapshot.
Particle-level views confirm that these structures arise from correlated microscopic motion, with individual velocities coherently aligned within the flow patterns (Fig.~\ref{fig:snapshot}g--i).

These structures are instantaneous realizations of a dynamically evolving state. The flow patterns reorganize once particles have been driven over a few particle diameters, that is, on the timescale $\tau_v \propto 1/V$, as identified above from the velocity autocorrelation functions (Fig.~\ref{fig:vv}).

Similar coherent structures were reported in Ref.~\cite{mo_rheological_2020} for the same driving protocol, where particles with large displacements accumulated over a driving distance of a few particle diameters form a band moving in a flocking pattern. Given the small system size studied there, we interpret this band as one of the correlated streams observed here, spanning the simulation box.

\subsection{Spatial correlations of the velocity field}

\label{cr}
\begin{figure}
    \centering
    \includegraphics{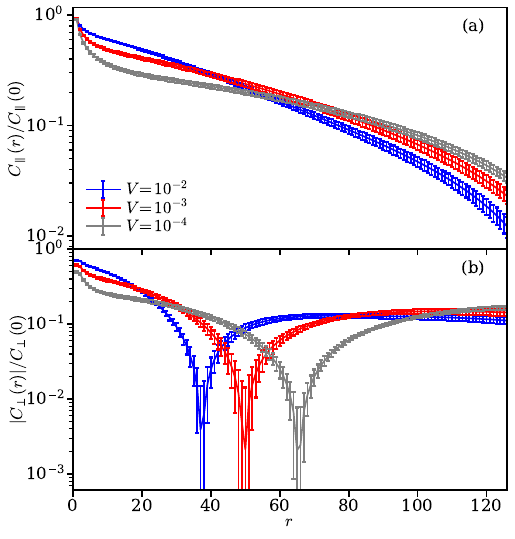}
    \caption{(a) Longitudinal and (b) transverse spatial velocity correlations for $N = 64 \times 10^3$ at different driving rates $V$. Error bars indicate the standard error over independent runs.}
    \label{fig:cr}
\end{figure}

To make the velocity correlations underlying the observed flow patterns quantitative, we compute equal-time spatial velocity correlations resolved into longitudinal ($\lambda = \parallel$) and transverse ($\lambda = \perp$) components~\cite{keta_emerging_2024}:
\begin{equation}
C_{\lambda}(r) = \frac{\sum_{i,j}v^{\lambda}_i v^{\lambda}_j\delta(r_{ij}-r)}{\sum_{i,j}\delta(r_{ij}-r)}\,,
\label{eq:Cr}
\end{equation}
where $v^{\lambda}_i$ denotes the projection of the particle velocity along or perpendicular to the unit separation vector $(\mathbf{r}_i-\mathbf{r}_j)/r_{ij}$. This decomposition is relative to the separation vector, unlike the one relative to $\hat{\mathbf{n}}_i$ used in Sec.~\ref{sec:single}. In Fig.~\ref{fig:cr}, each component is normalized by its zero-separation value, $C_\parallel(0)=C_\perp(0)=\langle|\mathbf{v}|^2\rangle/2$ for an isotropic two-dimensional system.

In a flow made of streams and vortices, $C_\parallel(r)$ captures the alignment of velocities along streams, while $C_\perp(r)$ becomes negative at distances comparable to the vortex size, where velocities on opposite sides of a vortex point in opposite directions. The total velocity correlation is $C(r) = C_{\parallel}(r) + C_{\perp}(r)$, and this decomposition disentangles the various geometric features revealed in the snapshots in  Fig.~\ref{fig:snapshot}.

The longitudinal correlation $C_\parallel(r)$ first drops over a few particle diameters, reflecting short-ranged and short-lived relaxations, and then decays approximately exponentially over a much larger distance (Fig.~\ref{fig:cr}a). Fitting this slow decay to $C_\parallel(r) \simeq A\,e^{-r/\xi_\parallel}$ gives $\xi_\parallel = 34$, $45$ and $66$ for $V = 10^{-2}$, $10^{-3}$ and $10^{-4}$ in our largest system. 

The transverse correlation $C_\perp(r)$ changes sign at a comparable distance, so both functions consistently show that the correlated flows grow as $V$ decreases, as also seen in Fig.~\ref{fig:snapshot}. In smaller systems we find systematically smaller values, and within the accessible sizes we could not reach a regime where these lengths stop changing with $N$. 

A similar behavior was reported for dense active fluids with infinitely persistent forces of fixed amplitude, where the correlation length grows proportionally to the linear system size over a broad range of sizes~\cite{Dutta2025, Shukla2025}. The values above should therefore be read as lower bounds of the intrinsic correlation length. Thus, while a finite $V$ introduces a well-defined timescale, the associated length scale quickly exceeds the size of our largest simulated system.

\subsection{Kinetic energy spectra in Fourier space}

\label{ek}

To further characterize the collective flow, we compute the kinetic energy spectrum
\begin{equation}
    E(k)=\frac{2\pi}{L^2} k \langle |\tilde{\mathbf{v}}(\mathbf{k})|^2 \rangle\,,
    \label{Eq:E(k)}
\end{equation}
with $k = |\mathbf{k}|$ and $\tilde{\mathbf{v}}(\mathbf{k})=\int{\rm d}^2\mathbf{r}\,\mathbf{v}(\mathbf{r})\exp{(-i\mathbf{k}\cdot \mathbf{r})}$ the Fourier transform of the velocity field $\mathbf{v}(\mathbf{r}) = \sum_i \mathbf{v}_i \delta(\mathbf{r} - \mathbf{r}_i)$.
The energy spectrum contains the same information as the real-space correlations, but provides a complementary view of the flows and allows comparison with active turbulence.

\begin{figure}
    \centering
    \includegraphics{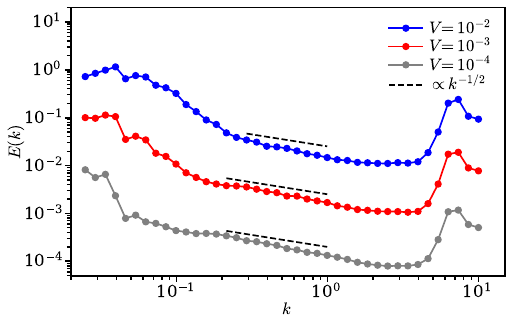}
    \caption{Kinetic energy spectra $E(k)$ for $N = 64000$ at different driving rates $V$, without rescaling. The dashed lines indicate the scaling $E(k) \sim k^{-1/2}$.}
    \label{fig:ek}
\end{figure}

The linear regime $E(k) \propto k$ expected at small $k$, once velocity correlations have decayed, requires system sizes much larger than the correlation length. It is only hinted at for $V = 10^{-2}$ and is not accessible at smaller $V$. Beyond the maximum, $E(k)$ first decays steeply over a narrow range of $k$. The regime is too affected by finite-size effects to reliably support a power law description. At larger $k$, however, the spectra display a much more robust regime, $E(k) \sim k^{-1/2}$, which extends over up to a decade and shifts to smaller $k$ as $V$ decreases (Fig.~\ref{fig:ek}). 

A similar scaling $k^{-1/2}$ was reported for the mesoscale flows of dense active particles with large but finite persistence times~\cite{keta_emerging_2024}, and for dense assemblies driven by non-reciprocal pairwise forces~\cite{KlamserBerthier2025}. This suggests that this $k^{-1/2}$ regime is a generic feature of the chaotic flows of locally driven dense disordered materials, and supports the view that the present system reaches the same turbulent state, even though the timescale is here set by the driving rate rather than by the persistence of the forces. Finally, the peak near $k \approx 2\pi/\sigma$ reflects the particle-based definition of the velocity field. It is not specific to the active turbulent flow, but simply reflects the local structure of the fluid at nearest neighbor distance.

\section{Conclusion}

\label{sec:conclusion}

We studied dense active matter with infinitely persistent forces, driven at a constant average velocity $V$ along the propulsion directions, following the protocol of Ref.~\cite{mo_rheological_2020}. The directions of the active forces never decorrelate, but $V$ introduces a finite timescale, $\sigma/V$, analogous to the inverse strain rate in sheared amorphous solids. The flow curve relating the average propulsion force to $V$ is of Herschel--Bulkley type, with a finite yield force $F_{\mathrm{c}}$ reached as $V \to 0$ and an exponent $n \simeq 1/2$, as found for a broad range of sheared materials~\cite{Bonn2017}. Since imposing the velocity or the force amplitude is equivalent in the thermodynamic limit~\cite{mo_rheological_2020}, imposing $V$ is a matter of convenience, which allows us to bring the system toward the flow threshold in a controlled way.

At the single-particle level, the velocity statistics reveal two contributions to the dynamics. The first consists of fast, localized plastic events. They produce fat tails in the velocity distribution, compatible at the smallest $V$ with the $v^{-3}$ decay expected from the $1/r$ elastic field of plastic events in two dimensions~\cite{PicardBocquet2004,Tanguy2006}, and they relax on a microscopic timescale that depends only weakly on $V$. Their weight increases strongly as $V$ decreases, and at the smallest driving rate they carry most of the velocity fluctuations. The second contribution is a slow collective flow, which decorrelates once particles have been driven over a few particle diameters, so that its lifetime scales as $1/V$. Beyond this time, each particle drifts along its own propulsion direction: the mean-squared displacement along $\hat{\mathbf{n}}_i$ becomes ballistic, the transverse one becomes diffusive, and the velocity autocorrelation function reaches a plateau $V^2/\langle|\mathbf{v}|^2\rangle=\chi\,V/F$. By analogy with sheared materials, the drift along $\hat{\mathbf{n}}_i$ plays the role of the affine displacement, while the transverse motion probes non-affine dynamics.

At the collective level, the velocity field organizes into interconnected streams and vortices whose spatial extent grows as $V$ decreases. These correlations emerge from the dynamics, since the propulsion directions develop only weak spatial correlations, without clear coarsening over the duration of our simulations. The decay of longitudinal velocity correlations and the sign change of transverse ones give consistent length scales, which grow from about 34 to 66 particle diameters as $V$ decreases from $10^{-2}$ to $10^{-4}$. These lengths keep increasing with system size and are therefore lower bounds of the intrinsic correlation length, as also found at fixed force~\cite{Dutta2025, Shukla2025}. The kinetic energy spectrum displays a robust $k^{-1/2}$ regime that extends to smaller wavevectors as $V$ decreases. The same scaling was found at large but finite persistence~\cite{keta_emerging_2024} and for non-reciprocal driving~\cite{KlamserBerthier2025}, which suggests a generic turbulent state of locally driven dense disordered materials. Approaching the flow threshold by decreasing $V$ therefore leads to the same active turbulent state as increasing the persistence time, with the driving rate controlling the lifetime of the correlated flows and, within the accessible system sizes, their spatial extension.

Overall, the analogy with sheared amorphous solids~\cite{berthier2025yielding} holds at several levels: a Herschel--Bulkley flow curve with $n \simeq 1/2$ and a force threshold, fat-tailed velocity distributions produced by plastic events, a drift playing the role of the affine displacement, and a reorganization of the flow after a fixed driving distance, much as after a fixed strain. The analogy also has clear limits. Because the driving is local and random, there is no macroscopic affine flow, and the non-affine motion organizes into streams and vortices with the spectral signature of active turbulence. Moreover, within the accessible system sizes, the correlation lengths do not saturate at finite $V$, unlike what is expected in sheared solids at finite strain rate~\cite{HeussingerChaudhuriBarrat2010}, a point we return to below.

Several questions remain worth pursuing. The first concerns length scales. Our largest systems cannot determine whether a nonzero $V$ eventually sets a finite correlation length, as expected by analogy with sheared solids at nonzero strain rate~\cite{HeussingerChaudhuriBarrat2010}, or whether the lengths keep growing with system size, as reported in simulations at fixed force amplitude above $F_{\mathrm{c}}$, which also correspond to a nonzero flow velocity $V$~\cite{Dutta2025, Shukla2025}. Answering this requires even larger simulated systems, and also observables that separate the slow collective flow from the plastic events. Correlations of particle displacements measured over a fixed driving distance, rather than of instantaneous velocities, would be a natural choice, analogous to non-affine displacements measured over a fixed strain interval in sheared solids~\cite{Tanguy2006, LemaitreCaroli2007, Heussinger2010}. The question is further complicated by a possible slow segregation of particles with similar propulsion directions~\cite{Dutta2025}, which makes the order of the limits of long times and large system sizes relevant.

A second set of questions concerns the flow threshold itself. The plastic events identified here call for a direct study of avalanche statistics in the steady state at nonzero $V$. Such a study would extend to nonzero driving rates the comparison between local and global driving performed in the quasistatic limit~\cite{morse_manning_2021, Agoritsas_2021, agoritsas_morse_2024}, and would help clarify whether the critical properties at the threshold are the same for local and global driving~\cite{villarroel_critical_2021, villarroel_avalanche_2024, relmucao-leiva_bridging_2025}. The threshold $F_{\mathrm{c}}$ also separates an arrested state from a chaotic flowing one. A similar transition driven by non-reciprocal forces was recently found to belong to the directed percolation universality class~\cite{KlamserBerthier2025}. It would be interesting to determine whether the onset of flow at $F_{\mathrm{c}}$ in persistent active matter falls in that same class, which requires simulations at fixed force amplitude, where the arrested state below $F_{\mathrm{c}}$ is also accessible. Finally, at the largest driving rate and system size, the homogeneous flowing state gives way to phase separation, as also observed at fixed force~\cite{Dutta2025}. How this density instability interacts with the correlated flows remains to be better characterized. 

\section*{Data Availability}

The data that support the findings of this study are openly available~\cite{zenododataset}.

\acknowledgments

We thank P. Chaudhuri and C. Dasgupta for useful discussions. L. B. acknowledges the support of the French Agence Nationale de la Recherche (ANR), under grants ANR-20-CE30-0031 (project THEMA) and ANR-24-CE30-0442 (project GLASSGO).

\bibliography{main.bib}

\end{document}